\documentclass[aps,twocolumn,pra,showpacs,superscriptaddress,10pt]{revtex4-1}
\usepackage{latexsym,amsmath,amssymb,amsfonts,graphicx,color,amsthm}
\usepackage{enumerate}
\usepackage{braket}
\usepackage{adjustbox}
\usepackage{footnote}
\usepackage[dvipsnames]{xcolor}
\usepackage{tipa}
\usepackage{textcomp}
\usepackage{fontenc}
\usepackage{mathrsfs}
\usepackage{color}
\usepackage{colortbl}
\usepackage{graphics,graphicx}
\usepackage{txfonts}
\usepackage{subcaption}
\usepackage{mathtools}
\usepackage{amssymb}
\usepackage{epsfig}
\usepackage{epstopdf}
\usepackage{dsfont}
\usepackage{todonotes}
\usepackage[colorlinks=true,linkcolor=blue,urlcolor=magenta,citecolor=red]{hyperref}
\usepackage[normalem]{ulem}

\begin{document}

\title{Generation of hyperentangled states and two-dimensional quantum walks using $J$- ($q$)- plates and polarization beamsplitters}
\author{P. A. Ameen Yasir}
\email{ameenyasir@iisc.ac.in}
\affiliation{Department of Instrumentation \& Applied Physics, Indian Institute of Science, Bengaluru 560012, India}
\author{C. M. Chandrashekar}
\email{chandracm@iisc.ac.in}
\affiliation{Department of Instrumentation \& Applied Physics, Indian Institute of Science, Bengaluru 560012, India}
\affiliation{The Institute of Mathematical Sciences, C. I. T. Campus, Taramani, Chennai 600113, India}
\affiliation{Homi Bhabha National Institute, Training School Complex, Anushakti Nagar, Mumbai 400094, India}

\begin{abstract}
A single photon can be made to entangle simultaneously in its different internal degrees of freedom\,(DoF) -- polarization, orbital angular momentum\,(OAM), and frequency -- as well as in its external DoF -- path. Such entanglement in multiple DoF is known as hyperentanglement and provide additional advantage for quantum information processing. We propose a  passive optical setup using $q$-plates and polarization beamsplitters to hyperentangle an incoming single photon in polarization, OAM, and path DoF. By mapping polarization DoF to a two-dimensional coin state, and path and OAM DoF to two spatial dimensions, $x$ and $y$, we present a scheme for realization of two-dimensional discrete-time quantum walk using only polarization beamsplitters and $q$-plates ensuing the generation of hyperentangled states. The amount of hyperentanglement generated is quantified by measuring the entanglement negativity between any two DoF.  We further show the hyperentanglement generation can be controlled by using an additional coin operation or by replacing the $q$-plate with  {one $J$-plate and 2 variable waveplates}. 
\end{abstract}


\maketitle

\section{Introduction} 
\label{in}

It is possible to entangle photons in more than one degree of freedom\,(DoF) such as polarization, time-energy, path, orbital angular momentum\,(OAM), and so on\,\cite{kwiat97,barreiro2005,zhao2019}. Such states are called hyperentangled states\,\cite{kwiat97}. Due to extension in the dimension of the Hilbert space of such paired photons, increase in the channel capacity has been demonstrated\,\cite{barreiro2008} and as a consequence hyperentanglement is poised to offer additional quantum advantage. Hyperentanglement in polarization and path DoF has been exploited in the context of entanglement purification protocols -- which has found applications in entanglement-based quantum key distribution\,\cite{sheng2010a,sheng2010b, sheng2010c, hu2021}.
Single photons can also be simultaneously entangled in polarization, path, and OAM DoF. While polarization and OAM correspond to internal DoF\,\cite{oneil2002} of the photon, path DoF corresponds to external DoF.  The amount of entanglement between these three DoF can be, for instance, generated and controlled using devices such as waveplates\,(both quarter and half-waveplates), polarization beamsplitters\,(PBS), and $q$-plates\,\cite{marrucci2006} or $J$-plates\,\cite{devlin2017}. 
Single photon entangled in these three DoF can also be thought of a quantum walker in higher dimensions.  For example, in a discrete-time quantum walk in two spatial dimensions, a coin degree of freedom can be mapped to polarization DoF and the two spatial dimensions can be mapped to path and OAM DoF. Therefore, controlled engineering of interactions between different DoF of single photon to generate and control hyperentanglement can be directly mapped to the controlled realization of higher dimensional quantum walks. 

Quantum walks, the quantum analog of classical random walks, are broadly classified into two categories\,: discrete-time quantum walk\,(DTQW) and continuous-time quantum walk\,(CTQW)\,\cite{andraca2012}. In the former case, the quantum coin dictates the direction in which the walker  moves and the walk evolves in the Hilbert space $\mathcal{H}_c \otimes \mathcal{H}_p$, where $\mathcal{H}_c$ denotes the Hilbert space corresponding to the coin space and $\mathcal{H}_p$ denotes the position space in which the walker moves. In the case of CTQW, no coin operation is necessary, and the state evolves only in the position Hilbert space. 

For one-dimensional DTQW the Hilbert space $\mathcal{H}_c$ is spanned by two-dimensional (2D) basis vectors $|H \rangle=[1,0]^T$ and $|V \rangle=[0,1]^T$. It can be noted that $|H \rangle$ and $|V \rangle$ denote Jones vectors corresponding to horizontal and vertical polarization states of photons, respectively. The Hilbert space $\mathcal{H}_p$ is spanned by the position basis $\{|x \rangle\}$, where $x \in \mathbb{Z}$. Each step of DTQW can be described using a composition of quantum coin operation $\hat{C}_{\boldsymbol \sigma}$ in SU(2),
\begin{align} \label{oq2}
\hat{C}_{\boldsymbol \sigma} = 
\begin{bmatrix}
e^{i\xi} \cos \theta & e^{i\zeta} \sin \theta \\
-e^{-i\zeta} \sin \theta & e^{-i\xi} \cos \theta
\end{bmatrix}
\end{align}
on $\mathcal{H}_c$, followed by a position shift operation, 
\begin{align} \label{oq1}
\hat{S} = \sum_{x=-\infty}^\infty [|H \rangle \langle H| \otimes |x-1 \rangle \langle x| + |V \rangle \langle V| \otimes |x+1 \rangle \langle x|]
\end{align}
on the combined Hilbert space\,\cite{chandru2008}.  After each step of walk operation the walker will evolve in superposition of position space entangling the two Hilbert spaces.  In an optical setting with polarization DoF, any $\hat{C}_{\boldsymbol \sigma}$ in SU(2) can be realized using two quarter waveplates and a half waveplate\,\cite{simon90} and $\hat{S}_x$ can be realized using PBS. In general, the 1D DTQW evolution after $n$ steps can be given by
\begin{align} \label{oq3}
|\Psi_n \rangle &= [\hat{S}_x (\hat{C}_{\boldsymbol \sigma} \otimes \mathds{1}_x)]^n |\Psi_{\rm in} \rangle \nonumber \\
&= \sum_{x=-\infty}^\infty [a_x^{(n)} |H \rangle + b_x^{(n)} |V \rangle] \otimes |x \rangle,
\end{align}
where $|\Psi_{\rm in} \rangle$ is the initial state, $\mathds{1}_x$ refers to identity operator in the position space, and $a_x^{(n)}$ and $b_x^{(n)}$ are normalized complex coefficients. The evolved state is evidently entangled in coin and spatial DoF.

For extension of DTQW to the 2D space, the Hilbert space will be a composition of $\mathcal{H}_{c} \otimes \mathcal{H}_{p_x} \otimes \mathcal{H}_{p_y}$, where $\mathcal{H}_{c}$ corresponds to the coin space and $\mathcal{H}_{p_x}$ and $\mathcal{H}_{p_y}$ refer to the Hilbert spaces corresponding to the position spaces in $x$ and $y$-directions, respectively. Since the state has to simultaneously evolve in both $x$ and $y$- spaces, it is natural to expect the use of 4-dimensional coin space and a corresponding coin operation. Two well-known examples of such coin choices are Grover coin and 4-dimensional discrete-Fourier transform coin\,\cite{tregenna2003}. However, it was shown that such 2D DTQW can as well be implemented using just 2D coin operation\,\cite{franco2011,chandru2010, chandru2013}. For instance, the Grover walk with an initial state $(\frac{1}{2}) (|0 \rangle -|1 \rangle -|2 \rangle +|3 \rangle) \otimes |x=0 \rangle \otimes |y=0 \rangle$ can be implemented using a two-state alternate walk -- in which a two dimensional coin operation is used and each step of walk is split into evolution in one dimension followed by an evolution in the other dimension.  It has also been demonstrated that the alternate walk can be implemented in the form of Pauli walk, where Pauli operators' bases are used as conditions in the shift operators and no coin operation is therefore necessary\,\cite{chandru2013}.

There has been a continued interest for efficient implementation of quantum walks\,({both in 1D and 2D spaces}) in various quantum systems. For example, in 1D, quantum walk has been realized using physical systems such as NMR\,\cite{ryan2005}, optical lattice\,\cite{karski2009}, linear optical devices\,\cite{schreiber2010,crespi2013, broome2010,zhang2007}, ion traps\,\cite{schmitz2009, zahringer2010}, and $q$-plates\,(single photons\,\cite{cardano2015} as well as bright classical light\,\cite{goyal2013,sephton2019}), to name a few. In 2D, the quantum walk has been realized using photonic waveguide arrays\,\cite{tang2018}, liquid-crystal devices\,\cite{errico2020}, etc. Various new schemes have been proposed for the realization of 1D quantum walk which include $q$-plates and waveplates\,\cite{zhang2010}, passive optical devices\,\cite{jeong2004,do2005}, cross-Kerr nonlinearity\,\cite{gao2019}.

Inspired by  the Pauli walk where different bases are used for evolution in different spatial dimensions, in this paper, we propose a passive optical setup -- using {$J$-plate + 2 variable waveplates}\,\cite{devlin2017} or $q$-plates\,\cite{marrucci2006} and polarization beamsplitters\,(PBS) -- to generate hyperentanglement in polarization, path, and OAM DoF of a single photon. Here, {$J$-plate + 2 variable waveplates} or $q$-plates will be used to control the OAM\,\cite{allen92} and polarization DoF, while PBS will be used to control the path DoF.  Upon evolution, we show that the photon will be hyperentangled in these three DoF. This setup can also effectively simulate  a 2D modified form of Pauli walk in OAM and position DoF where coin operation is not required.  Due to basis change that $J$- ($q-$) plate and PBS introduce, the effect of coin operation in the path dimension is absorbed into the $J$- ($q$-) plate and the effect of coin operation  in OAM dimension is absorbed into PBS. By mapping the path and OAM DoF to $x$ and $y$-dimensions we can recover the DTQW in two dimensional  position space. 

This paper is organized as follows. In Section\,\ref{tq} we briefly review schemes for realizing 2D DTQW such as Pauli and alternate walks and explain how the evolved state is hyperentangled in the associated Hilbert spaces.  In Section\,\ref{hy} we propose a passive optical setup to hyperentangle the incoming single photon in the three DoF\,(polarization, path, and OAM). This hyperentanglement is quantified by measuring the entanglement negativity between any two of the three DoF. We then present our numerical results by simulating the two-dimensional modified Pauli walk -- which does not require an explicit coin operator. Finally, in Section\,\ref{co} we conclude with some remarks.

\section{Two-dimensional quantum walk} 
\label{tq}

In this section we show the equivalence between the alternate and Pauli walks for any arbitrary choice of coin operator in SU(2). We also propose modified Pauli walk and discuss its implementation in optical setting using {$J$-plate + 2 variable waveplates} and PBS for a particular choice of SU(2) parameters. To quantify hyperentanglement, we use entanglement negativity which measures entanglement between any two of three DoF.

{\noindent \bf Mathematical framework\,:} Quantum walk in 2D can be implemented using 2D coin operator and shift operators in $x$ and $y$-directions\,\citep{chandru2010,chandru2013,franco2011}. We define coin operator as $\hat{C}_{\boldsymbol \sigma}$\,[Eq.\,(\ref{oq2})], and shift operators can be defined as
\begin{align} 
\hat{S}_{x} &= \sum_{x=-\infty}^\infty [|H \rangle \langle H| \otimes |x-1 \rangle \langle x| \otimes \mathds{1}_{y} \nonumber \\ 
&\,\,\,+ |V \rangle \langle V| \otimes |x+1 \rangle \langle x| \otimes \mathds{1}_{y}], \label{tq2a} \\
\hat{S}_{y} &= \sum_{y=-\infty}^\infty [|H \rangle \langle H| \otimes \mathds{1}_{x} \otimes |y-1 \rangle \langle y| \nonumber \\ 
&\,\,\,+ |V \rangle \langle V| \otimes \mathds{1}_{x} \otimes |y+1 \rangle \langle y|], \label{tq2b} 
\end{align}
where $\mathds{1}_x$ and $\mathds{1}_y$ are identity operators in $x$ and $y$ spaces, respectively. If $|\Psi_{\rm in} \rangle$ represents initial state, the evolution operator $\hat{\mathcal{O}}$ corresponding to {\it alternate walk} is\,\citep{chandru2013}
\begin{align} \label{tq3}
|\Psi_1 \rangle = \hat{S}_x [\hat{C}_{\boldsymbol \sigma} \otimes \mathds{1}_{xy}] \hat{S}_y [C_{\boldsymbol \sigma}^\dagger \otimes \mathds{1}_{xy}] |\Psi_{\rm in} \rangle = \hat{\mathcal{O}} |\Psi_{\rm in} \rangle,
\end{align}
where $\mathds{1}_{xy}=\mathds{1}_x \otimes \mathds{1}_y$. This alternate walk evolution operator can also be implemented using just two shift operators, $\hat{S}_x$ and $\hat{S}_{\boldsymbol \sigma}$, where
\begin{align} \label{tq4a}
\hat{S}_{\boldsymbol \sigma} &= [\hat{C}_{\boldsymbol \sigma} \otimes \mathds{1}_{xy}] \hat{S}_y [\hat{C}_{\boldsymbol \sigma}^\dagger \otimes \mathds{1}_{xy}] \nonumber \\
&= \sum_{y} [|u_1 \rangle \langle u_1| \otimes \mathds{1}_{x} \otimes |y-1 \rangle \langle y| \nonumber \\
&\,\,\,+ |u_2 \rangle \langle u_2| \otimes \mathds{1}_{x} \otimes |y+1 \rangle \langle y|].
\end{align}
The states $|u_1 \rangle$ and $|u_2 \rangle$ denote the first and second column vectors of $\hat{C}_{\boldsymbol \sigma}$\,[see Eq.\,(\ref{oq2})], respectively. In any physical system with provision to directly realize $\hat{S}_x$ and $\hat{S}_{\boldsymbol \sigma}$, without explicit use of coin operation,  we can realize a 2D DTQW. When $\{|u_1 \rangle, |u_2 \rangle\}$ are the eigenvectors of the Pauli matrices, 
\begin{align} \label{tq4b}
\sigma_1 = \begin{bmatrix}
0 & 1 \\
1 & 0
\end{bmatrix} \,\,\,{\rm or} \,\,\,
\sigma_2 = \begin{bmatrix}
0 & -i \\
i & 0
\end{bmatrix},
\end{align}
the evolution operator $\hat{S}_x \hat{S}_{\boldsymbol \sigma}$ readily implements the {\it Pauli walk}\,\cite{chandru2013}. Therefore, the operator $\hat{S}_x \hat{S}_{\boldsymbol \sigma}$ can be thought of as a {\it generalized Pauli walk} and its evolution can be given by
\begin{align} \label{tq5}
|\Psi_1 \rangle = \hat{S}_x \hat{S}_{\boldsymbol \sigma} |\Psi_{\rm in} \rangle = \hat{\mathcal{O}} |\Psi_{\rm in} \rangle.
\end{align}

\begin{figure*}[htbp]
\centering
\includegraphics[scale=0.27]{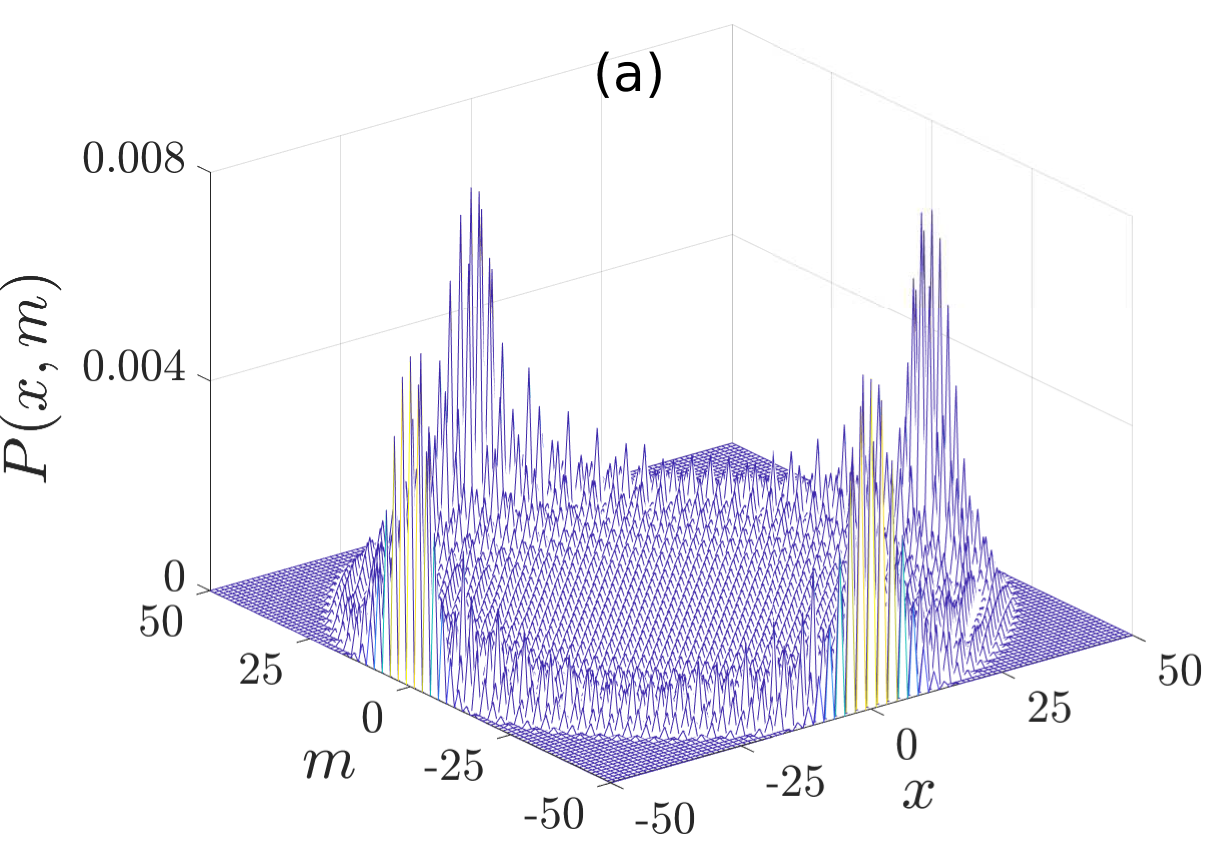}~~\includegraphics[scale=0.27]{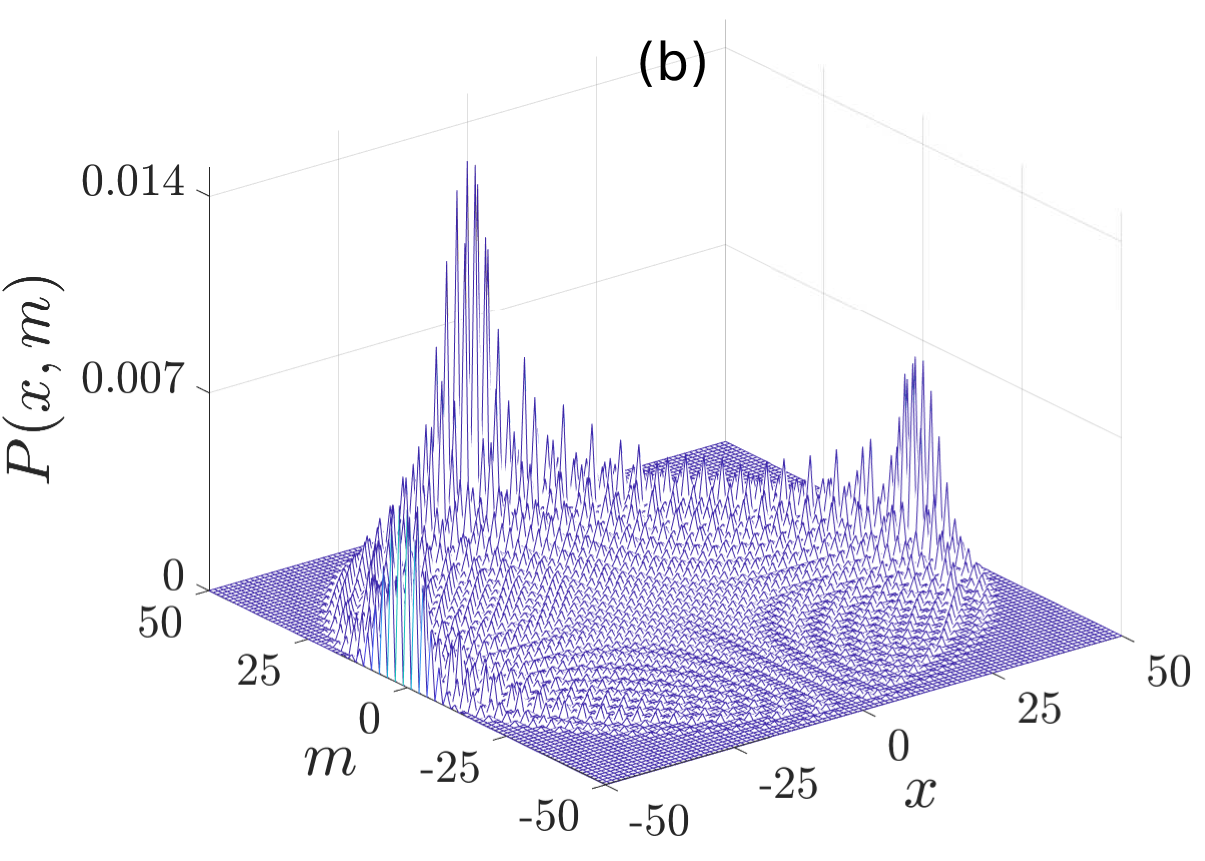}~~\includegraphics[scale=0.27]{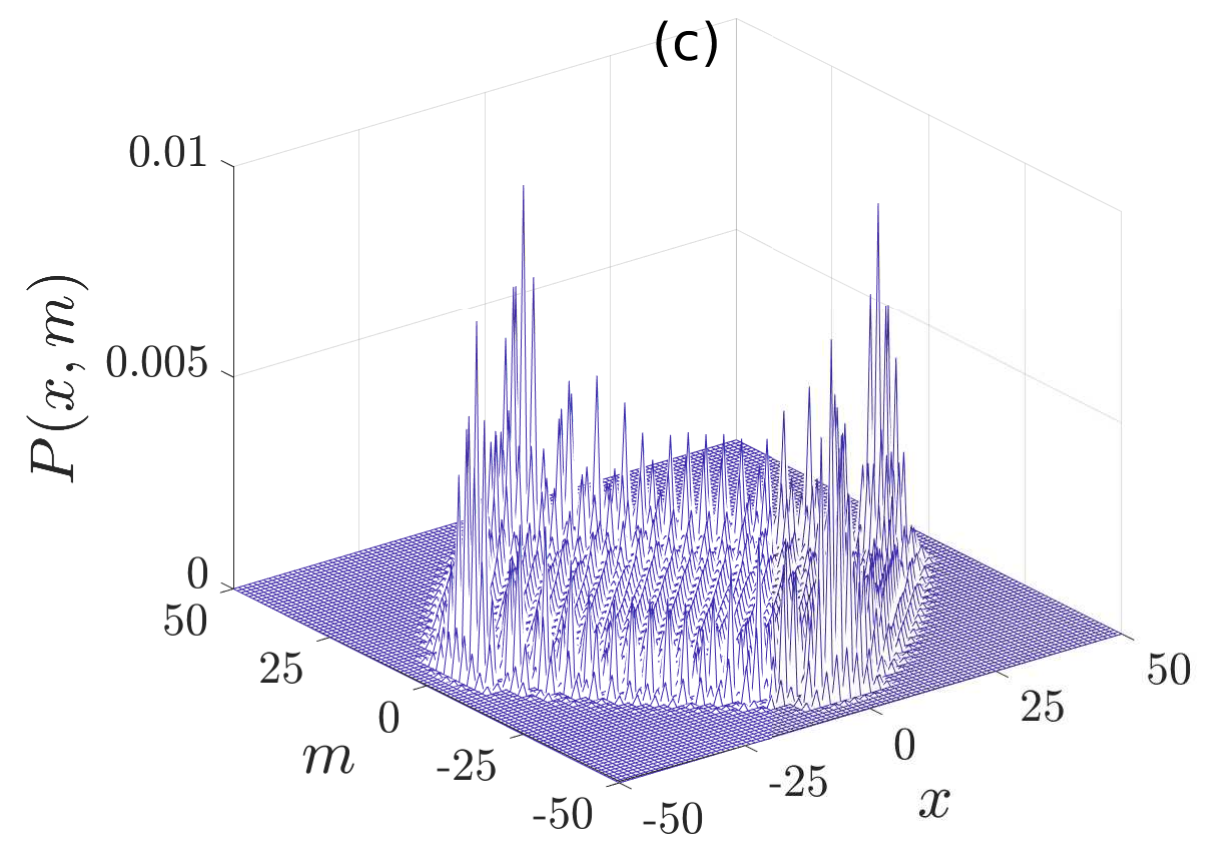}
\caption{Probability distribution\,($P(x,m)$) of the 2D DTQW without explicit coin operation in position and OAM DoF -- beginning with an initial state $[(|H \rangle+ |V \rangle)/\sqrt{2}] \otimes |x=0 \rangle \otimes |y=0 \rangle$ -- after 50 steps. (a) Modified Pauli walk using $q$-plates and PBS\,[see Eq.\,(\ref{tq14})]. Note that the probability distribution is identical for the Pauli walk\,[see Eqs.\,(\ref{tq5}) and (\ref{tq15})] which is realized using $q$-plates, PBS, and HWP. (b) Modified Pauli walk using {$J$-plate + 2 variable waveplates} and PBS for the choice $|u_1 \rangle=[1,-1]^T/\sqrt{2}$ and $|u_2 \rangle=[1,1]^T/\sqrt{2}$ in Eq.\,(\ref{tq14}). (c) Modified Pauli walk using {$J$-plate + 2 variable waveplates} and PBS for the choice $|u_1 \rangle=[1,\sqrt{3} i]^T/2$ and $|u_2 \rangle=[\sqrt{3}, -i ]^T/2$ in Eq.\,(\ref{tq14}).}
\label{prob}
\end{figure*}

Shift operators $\hat{S}_x$ and $\hat{S}_y$ given in Eqs.\,(\ref{tq2a}) and (\ref{tq2b}) shifts the position vector without changing the coin state vectors. However, we can also define a {\it modified shift operator} which induces  a flip in the coin state vector along with the shift in position vector. For example, we can define
\begin{align} \label{tq6a}
\hat{S}'_{y} = [\sigma_1 \otimes \mathds{1}_{xy}] \hat{S}_y &= \sum_{y=-\infty}^\infty [|V \rangle \langle H| \otimes \mathds{1}_{x} \otimes |y-1 \rangle \langle y| \nonumber \\ 
&\,\,\,+ |H \rangle \langle V| \otimes \mathds{1}_{x} \otimes |y+1 \rangle \langle y|],
\end{align}
and $\hat{S}'_x$ defined likewise. Because of the bit-flip symmetry\,\cite{chandru2007}, it can be shown that both $\hat{\mathcal{O}}$ and 
\begin{align} \label{tq6b}
\hat{\mathcal{O}}' \equiv \hat{S}_x [\hat{C}_{\boldsymbol \sigma} \otimes \mathds{1}_{xy}] \hat{S}'_y [C_{\boldsymbol \sigma}^\dagger \otimes \mathds{1}_{xy}] |\Psi_{\rm in} \rangle
\end{align}
lead to the equivalent evolution of the initial state $|\Psi_{\rm in} \rangle$ in Eq.\,(\ref{tq3}). {By equivalent evolution, we mean that both $\hat{\mathcal{O}}$ and $\hat{\mathcal{O}}'$ lead to the same position probability distribution.} Thus, we can also define {\it modified versions} of both alternate and generalized Pauli walks as
\begin{align}
|\Psi_1 \rangle &= \hat{S}_x [\hat{C}_{\boldsymbol \sigma} \otimes \mathds{1}_{xy}] \hat{S}'_y [\hat{C}_{\boldsymbol \sigma}^\dagger \otimes \mathds{1}_{xy}] |\Psi_{\rm in} \rangle, \label{tq7a} \\
{\rm and} \,\,\,
|\Psi_1 \rangle &= \hat{S}_x \hat{S}'_{\boldsymbol \sigma} |\Psi_{\rm in} \rangle, \label{tq7b}
\end{align}
respectively, with
\begin{align} \label{tq8}
\hat{S}'_{\boldsymbol \sigma} &= [\hat{C}_{\boldsymbol \sigma} \otimes \mathds{1}_{xy}] \hat{S}'_y [\hat{C}_{\boldsymbol \sigma}^\dagger \otimes \mathds{1}_{xy}] \nonumber \\
&= \sum_{y} [|u_2 \rangle \langle u_1| \otimes \mathds{1}_{x} \otimes |y-1 \rangle \langle y| \nonumber \\
&\,\,\,+ |u_1 \rangle \langle u_2| \otimes \mathds{1}_{x} \otimes |y+1 \rangle \langle y|].
\end{align}
If we begin with the initial state
\begin{align} \label{tq9}
|\Psi_{\rm in} \rangle = (\cos \alpha |H \rangle + e^{i\beta} \sin \alpha |V \rangle) \otimes |x=0 \rangle \otimes |y=0 \rangle,
\end{align}
then, after $n$ steps, the state will be of the form
\begin{align} \label{tq10}
|\Psi_n \rangle &= (\hat{S}_x \hat{S}'_{\boldsymbol \sigma})^n |\Psi_{\rm in} \rangle \nonumber \\ 
&= \sum_{x,y} \left[ (a_{x,y}^{(n)} |H \rangle + b_{x,y}^{(n)} |V \rangle) \otimes |x \rangle \otimes |y \rangle \right],
\end{align}
where $a_{x,y}^{(n)}$ and $b_{x,y}^{(n)}$ are normalized complex coefficients. The {recurrence} relations between $a_{x,y}^{(n)}$ and $b_{x,y}^{(n)}$ are
\begin{align} \label{tq11a}
a_{x,y}^{(n)} &= a_{x+1,y+1}^{(n-1)} (e^{i\zeta} \sin \theta) (e^{-i\xi} \cos \theta) \nonumber \\ 
&\,\,\,+ a_{x+1,y-1}^{(n-1)} (e^{i\xi} \cos \theta) (e^{-i\zeta} \sin \theta) \nonumber \\ 
&\,\,\,+ b_{x-1,y+1}^{(n-1)} (e^{-i\xi} \cos \theta) (e^{-i\xi} \cos \theta) \nonumber \\
&\,\,\,+ b_{x-1,y-1}^{(n-1)} (-e^{-i\zeta} \sin \theta) (e^{-i\zeta} \sin \theta),  
\end{align}
and
\begin{align} \label{tq11b}
b_{x,y}^{(n)} &= a_{x+1,y+1}^{(n-1)} (e^{i\zeta} \sin \theta) (-e^{i\zeta} \sin \theta) \nonumber \\
&\,\,\,+ a_{x+1,y-1}^{(n-1)} (e^{i\xi} \cos \theta) (e^{i\xi} \cos \theta) \nonumber \\ 
&\,\,\,+ b_{x-1,y+1}^{(n-1)} (e^{-i\xi} \cos \theta) (-e^{i\zeta} \sin \theta) \nonumber \\
&\,\,\,+ b_{x-1,y-1}^{(n-1)} (-e^{-i\zeta} \sin \theta) (e^{i\xi} \cos \theta). 
\end{align}
The above described state evolution after $n$ steps is in superposition of the tensor products of the three Hilbert spaces, namely coin Hilbert space $\mathcal{H}_c$, and two position Hilbert spaces $\mathcal{H}_{p_x}$ and $\mathcal{H}_{p_y}$ associated with the dynamics. The interwinding coefficients of the state vectors after evolution clearly indicate that the Eq.\,(\ref{tq10}) is {\it hyperentangled}\,\cite{kwiat97}.

{\noindent \bf Optical realization\,:} It is possible to realize the shift operators in Eq.\,(\ref{tq7b}) without an explicit coin operation using passive optical devices PBS, {$J$-plates and variable waveplates} on polarization and OAM DoF, respectively. Operator $\hat{S}_x$\,($\hat{S}_{x,\rm pos}$ from now on, with `pos' referring to the position DoF) can be readily realized using the PBS -- which reflect horizontal polarization and transmit vertical polarization. Operator $\hat{S}'_{\sigma}$\, ($\hat{S}'_{\boldsymbol \sigma,{\rm OAM}}$) can be realized using a {$J$-plate + 2 variable waveplates}. 

{To understand the action of $J$-plate let us consider the light field $\psi(r,\phi;z)$ propagating in the $z$-direction, where $r=\sqrt{x'^2+y'^2}$ and $\phi=\tan^{-1} (y'/x')$ with $(x',y')$ being the coordinates in the transverse plane.  The light field $\psi(r,\phi;z)$ carrying an OAM of $m\hbar$ per photon\,\cite{allen92} can be written as
\begin{align} \label{oam}
\psi(r,\phi;z) \propto A(r;z) \exp (im\phi),
\end{align}
where $A(r;z)$ denotes the amplitude profile and $\exp (im\phi)$ denotes the phase profile. If this light field in the polarization state $|u_1 \rangle$ passes through a {$J$-plate + waveplates combination\,(see Appendix\,\ref{ap})} represented by a Jones matrix
\begin{align} \label{tq12a}
J(\phi) = e^{-i\phi} |u_2 \rangle \langle u_1| + e^{i\phi} |u_1 \rangle \langle u_2|,
\end{align}   
then its polarization vector\,(or Jones vector) will change to $|u_2 \rangle$ and its phase profile will transform as $\exp {[i(m-1)\phi]}$, where $|u_2 \rangle$ is the Jones vector orthogonal to $|u_1 \rangle$. Likewise, the phase profile of the light field $\psi(r,\phi;z)$ in the polarization state $|u_2 \rangle$ will be transformed to $\exp {[i(m+1)\phi]}$ by the action of {$J$-plate + waveplates combination}, while the polarization state being changed to $|u_1 \rangle$. Therefore, we find that the OAM of the light field has been reduced by $\hbar$ per photon in the former case, whereas it has been increased by $\hbar$ per photon in the latter case. 

{Now let us consider a single photon carrying an OAM of $m\hbar$ per photon in some polarization state. The {$J$-plate + waveplates combination} can decrease (increase) the OAM of the incoming photon with Jones vector $|u_1 \rangle$ ($|u_2 \rangle$) by $\hbar$ per photon while simultaneously transforming the Jones vector of the photon to $|u_2 \rangle$\,($|u_1 \rangle$). In other words, {the $J$-plate + waveplates combination} changes the OAM of the incoming single photon conditioned over the polarization states $\{|u_1 \rangle, |u_2 \rangle\}$. Since $|u_1 \rangle$ and $|u_2 \rangle$ are themselves functions of $(\xi,\zeta,\theta)$\,[Eqs.\,(\ref{oq2}) and (\ref{tq4a})], $J(\phi)$ in Eq.\,(\ref{tq12a}) can also be written as 
\begin{align} \label{tq12b}
J(\phi) \equiv J(\phi,\xi,\zeta,\theta).
\end{align}
Thus, the shift operator realizing this transformation will be\,[cf. Eq.\,(\ref{tq8})],
\begin{align} \label{tq13}
\hat{S}'_{\boldsymbol \sigma,{\rm OAM}} &= \sum_{m} [|u_2 \rangle \langle u_1| \otimes \mathds{1}_{x,{\rm pos}} \otimes |m-1 \rangle \langle m| \nonumber \\
&\,\,\,+ |u_1 \rangle \langle u_2| \otimes \mathds{1}_{x,{\rm pos}} \otimes |m+1 \rangle \langle m|].
\end{align}
For the special case when $|u_1 \rangle=|R \rangle=[1,-i]^T/\sqrt{2}$ and $|u_2 \rangle=|L \rangle=[1,i]^T/\sqrt{2}$, or equivalently for the choice $J(\phi,0,-\pi/2,\pi/4)$, $\hat{S}'_{\boldsymbol \sigma,{\rm OAM}}$ is realized using a $q$-plate\,\cite{marrucci2006}. With these, we find that a single photon in the initial state $|\Psi_{\rm in} \rangle$, under the action of PBS and {$J$-plate + waveplates combinations}, will evolve as
\begin{align} \label{tq14}
|\Psi_n \rangle &= (\hat{S}_{x,{\rm pos}} \hat{S}'_{\boldsymbol \sigma, {\rm OAM}})^n |\Psi_{\rm in} \rangle \nonumber \\ 
&= \sum_{x,m} \left[ (a_{x,m}^{(n)} |H \rangle + b_{x,m}^{(n)} |V \rangle) \otimes |x \rangle \otimes |m \rangle \right],
\end{align}
where the normalized complex coefficients are iteratively related as in Eqs.\,(\ref{tq11a}) and (\ref{tq11b}) with the position label $y$ being replaced by the OAM label $m$. 

We can also realize 
\begin{align} \label{tq15}
\hat{S}_{\boldsymbol \sigma,{\rm OAM}} &= \sum_{m} [|u_1 \rangle \langle u_1| \otimes \mathds{1}_{x,{\rm pos}} \otimes |m-1 \rangle \langle m| \nonumber \\
&\,\,\,+ |u_2 \rangle \langle u_2| \otimes \mathds{1}_{x,{\rm pos}} \otimes |m+1 \rangle \langle m|]
\end{align}
[see Eq.\,(\ref{tq4a})] using {the $J$-plate + 2 variable waveplates\,(see Appendix\,\ref{ap})} with Jones matrix
\begin{align} \label{tq16}
\tilde{J}(\phi) &= e^{-i\phi} |u_1 \rangle \langle u_1| + e^{i\phi} |u_2 \rangle \langle u_2| \nonumber \\
&\equiv \tilde{J}(\phi,\xi,\zeta,\theta).
\end{align}
However, in order to realize $\hat{S}_{\boldsymbol \sigma, {\rm OAM}}$ with $|u_1 \rangle=|R \rangle$ and $|u_2 \rangle=|L \rangle$ using a $q$-plate instead of the {$J$-plate + QHQ combination}, we will require an additional half-waveplate\,(HWP). 

\begin{figure}[htbp]
\centering
\includegraphics[scale=0.3]{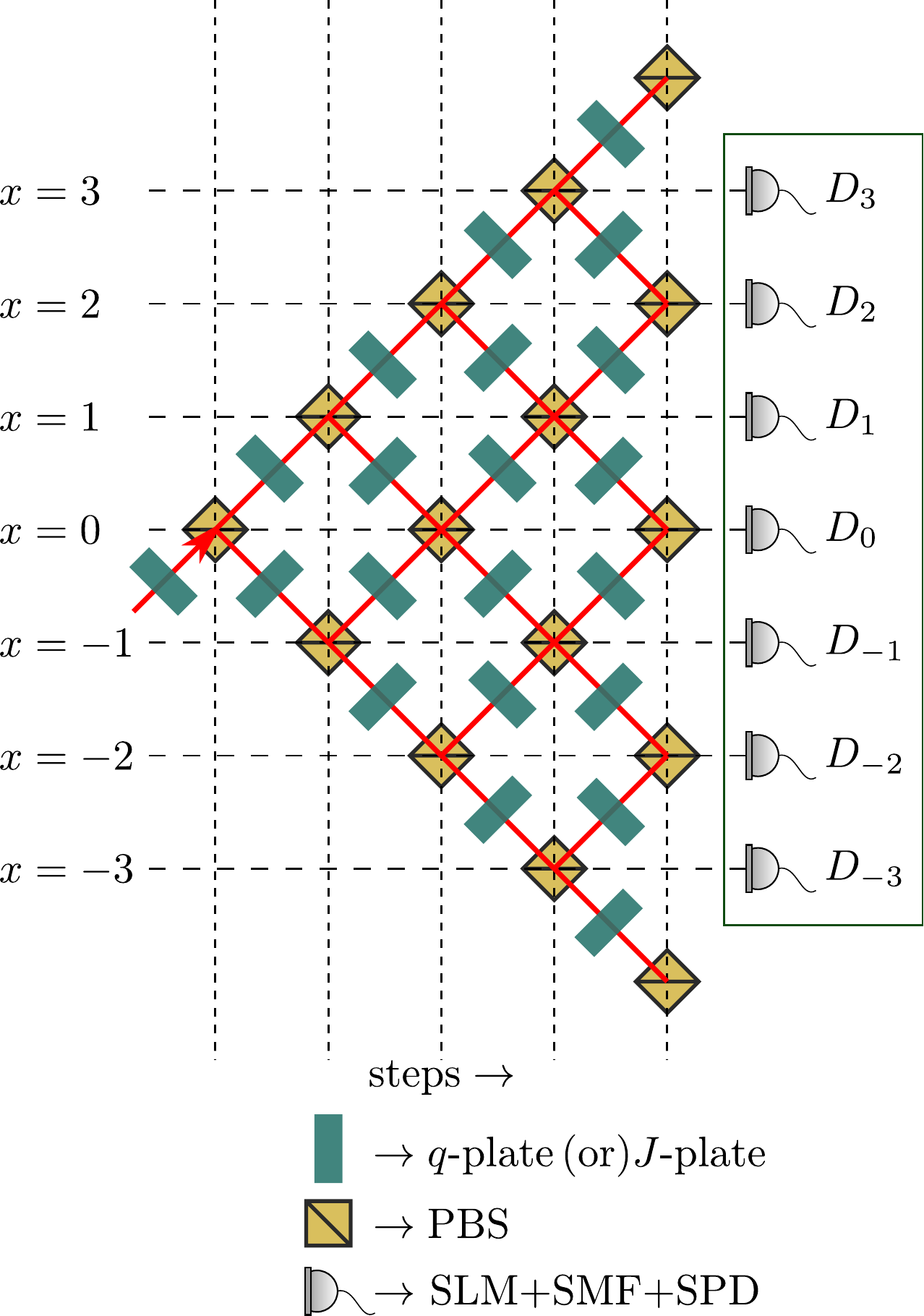}
\caption{Optical implementation of Pauli and modified Pauli walks. Both Pauli and modified Pauli walks with any orthogonal set of vectors $\{|u_1 \rangle, |u_2 \rangle\}$ can be realized using  {$J$-plate + waveplates combination} and PBS\,[see Eqs.\,(\ref{tq5}), (\ref{tq14}), and (\ref{tq15})]. When $\{|u_1 \rangle, |u_2 \rangle\}=\{|R \rangle, |L \rangle\}$, every {$J$-plate + waveplates combination} can be replaced with a $q$-plate\,($q$-plate and HWP) to realize the modified Pauli walk\,(Pauli walk). $D_i$ denotes a detector unit placed at the position $x=i$. Each detector unit consists of a spatial light modulator\,(SLM), a single mode fiber\,(SMF), and a single photon detector\,(SPD). A single photon is sent through a $q$-plate\,(or  {$J$-plate + waveplates combination}) to the PBS placed at $x=0$\,(shown with an arrow). Here, {$J$-plate + waveplates combination} implements either one of the shift operators $\hat{S}'_{\boldsymbol \sigma,{\rm OAM}}$\,[Eq.\,(\ref{tq13})] or $\hat{S}_{\boldsymbol \sigma,{\rm OAM}}$\,[Eq.\,(\ref{tq15})], and PBS implements the shift operator $\hat{S}_{x,{\rm pos}}$\,[see Eq.\,(\ref{tq14})]. Evidently, no explicit {2D} coin operation is necessary to implement these type of quantum walks.}
\label{setup}
\end{figure}

The probability distribution of both Pauli and modified Pauli walks for $n=50$ steps beginning with an initial state $|\Psi_{\rm in} \rangle=[(|H \rangle+ |V \rangle)/\sqrt{2}] \otimes |x=0 \rangle \otimes |m=0 \rangle$ [that is, by substituting $\alpha=\pi/4$ and $\beta=0$ in Eq.\,(\ref{tq9})] has been shown in Fig.\,\ref{prob}. In (a) we have considered modified Pauli walk using $q$-plates and PBS\,[see Eq.\,(\ref{tq14})]. Owing to bit-flip symmetry\,\cite{chandru2007}, the probability distribution for the Pauli walk -- realized using $q$-plates, PBS, and HWP -- will also be identical to that of (a)\,[Eqs.\,(\ref{tq5}) and (\ref{tq15})]. In (b) and (c) we have considered modified Pauli walk -- realized using {$J$-plate + waveplates combinations} and PBS. The orthogonal state vectors $\{|u_1 \rangle, |u_2 \rangle\}$ for (b) and (c) were chosen to be 
\begin{align*}
& \{ [1,-1]^T/\sqrt{2}, [1,1]^T/\sqrt{2} \}, \,\,\,{\rm and} \\
& \{ [1,\sqrt{3} i]^T/2, [\sqrt{3},-i]^T/2 \},
\end{align*} 
respectively.

\section{Generation of Hyperentanglement} 
\label{hy}

\begin{figure*}[htbp]
\centering
\includegraphics[scale=0.4]{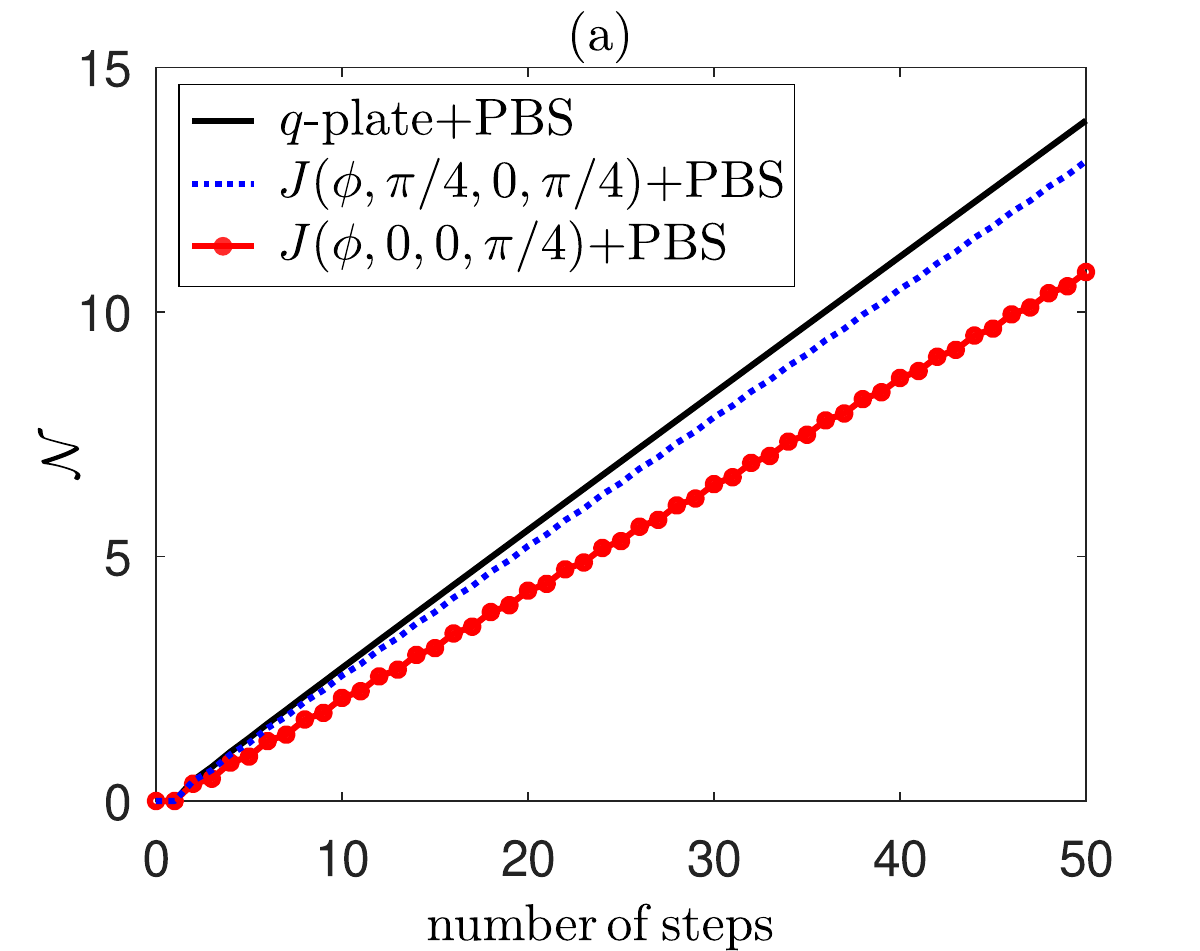}~~\includegraphics[scale=0.4]{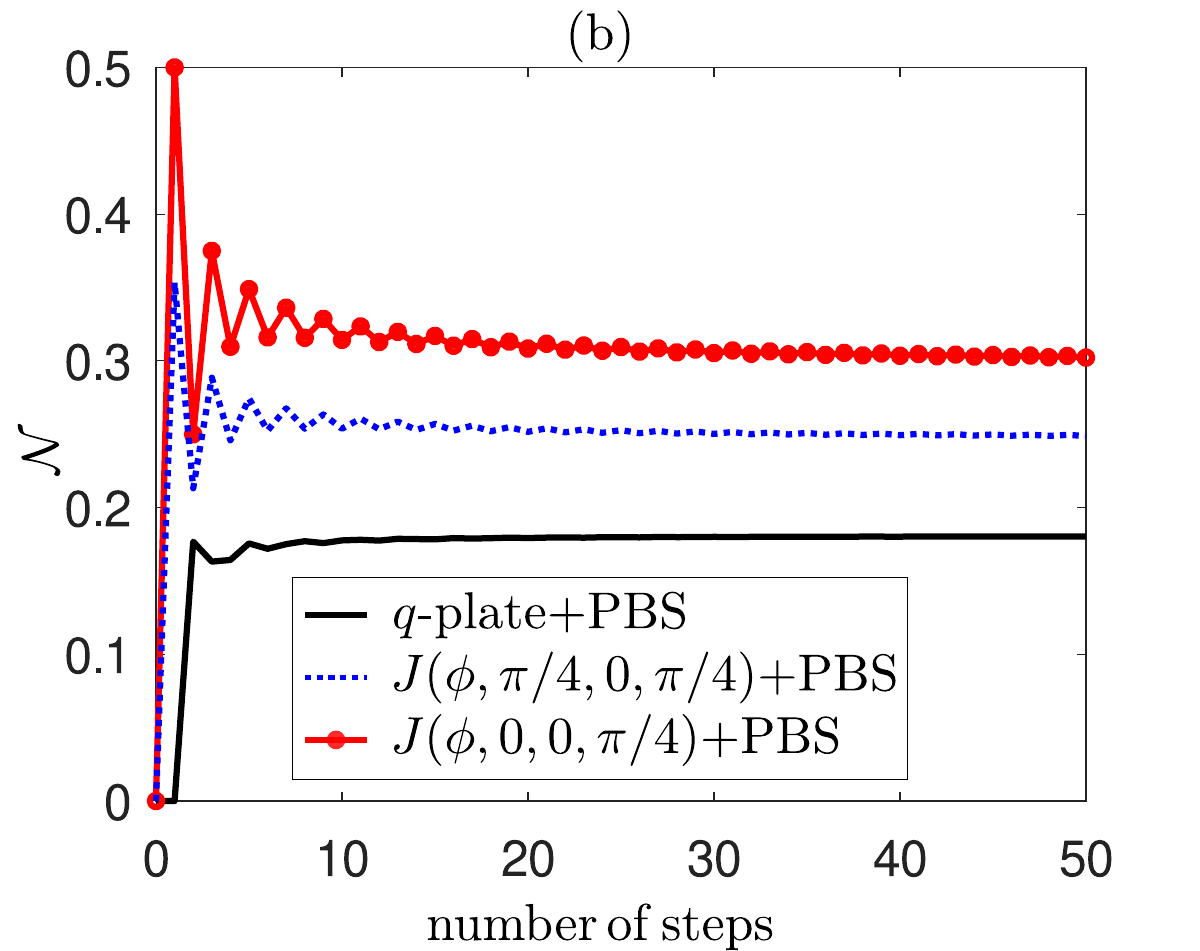}
\caption{Entanglement negativity $\mathcal{N}$\,[see Eq.\,(\ref{ne})] plotted against increasing number of steps between two DoF for various $J$-plate parameters $(\xi,\zeta,\theta)$\,[see (\ref{tq12b})]. Here, $J(\phi,0,-\pi/2,\pi/4)$ represents a $q$-plate, and the initial state is $[(|H \rangle+|V \rangle)/\sqrt{2}] \otimes |x=0 \rangle \otimes |y=0 \rangle$. (a) $\mathcal{N}$ between path and OAM DoF. (b) $\mathcal{N}$ between polarization and OAM DoF. Note that $\mathcal{N}$ between polarization and path DoF is identical as that of (b).}
\label{steps}
\end{figure*}


In this section we present an outline of the optical setup which can hyperentangle the incoming single photon in polarization, path, and OAM DoF and realize 2D DTQW. The hyperentanglement between the three DoF involved in the dynamics is quantified using entanglement negativity between the combination of the Hilbert spaces. 

In Fig.\,\ref{setup} we present the schematic representation of the setup for optical implementation of both Pauli and modified Pauli walks. In the case of modified Pauli walk, the shift operator $\hat{S}_{x,{\rm pos}}$ is realized using a PBS, $\hat{S}'_{\boldsymbol \sigma, {\rm OAM}}$ is realized using {the $J$-plate + waveplates combination} for any given orthogonal set of vectors $\{|u_1 \rangle, |u_2 \rangle\}$\,[see Eq.\,(\ref{tq14})]. To realize the Pauli walk, we just have to replace $\hat{S}'_{\boldsymbol \sigma, {\rm OAM}}$ in Eq.\,(\ref{tq14}) with $\hat{S}_{\boldsymbol \sigma, {\rm OAM}}$ in Eq.\,(\ref{tq15}). Clearly, $\hat{S}_{\boldsymbol \sigma, {\rm OAM}}$ is also realized using {the $J$-plate + waveplates combination} with the Jones matrix given in Eq.\,(\ref{tq16}). When $\{|u_1 \rangle, |u_2 \rangle\}=\{|R \rangle, |L \rangle\}$, every {$J$-plate + waveplates combination} can be replaced with a $q$-plate in the modified Pauli walk. On the other hand, every {$J$-plate + waveplates combination} has to be replaced with a $q$-plate and a HWP to realize the Pauli walk. 

While $\hat{S}_{x,{\rm pos}}$ controls the path DoF of the single photon, $\hat{S}'_{\boldsymbol \sigma, {\rm OAM}}$\,(or $\hat{S}_{\boldsymbol \sigma, {\rm OAM}}$) controls both polarization and OAM DoF of the same. Therefore, we don't need to explicitly use a coin operation to control the polarization DoF. This setup requires $n(n+1)/2$ PBS, {$n(n-1)+1$ $J$-plates, and $2[n(n-1)+1]$ variable waveplates to realize the Pauli walk for $n$ steps. In the case of the modified Pauli walk with the same number of steps, we will require an additional $n(n-1)+1$ half waveplates}. Here we have two remarks to make. First, the number of PBS and $J$-plates required to implement this type of quantum walk will scale quadratically with the number of steps. Second, when $\{|u_1 \rangle, |u_2 \rangle\}=\{|H \rangle, |V \rangle\}$, the evolved state is localized at the center and no spread is therefore observed. 

The single photon, after $n$-steps, would have evolved in superposition of position and OAM space. Upon measurement it will collapse at any one of the detector units $D_0$, $D_1$, $D_{-1}$, $\ldots$, placed as shown in the Fig.\,\ref{setup}. Each detector unit contains a spatial light modulator\,(SLM), a single mode fiber\,(SMF), and a single photon detector\,(SPD). The measurement of the OAM DoF requires all three of these components\,\cite{zhang2010,cardano2015}, whereas the measurement of the path DoF requires just a SPD\,\cite{do2005}. {To realize the 2D DTQW, we note that the Hilbert space corresponding to the photon's path represents one spatial dimension, $x$-axis. Since the photon at each position $x$ can end up with an OAM value of $y\hbar$ per photon, it represents the second spatial dimension, $y$-axis}.

After $n$ steps, the single photon will be entangled in polarization, path, and OAM DoF. To quantify the amount of entanglement between any two DoF, we adopt a measure known as the {\it entanglement negativity}\,\cite{vidal2002}. Here, we first partial trace out the density matrix $|\Psi_n \rangle \langle \Psi_n|$ with respect to the third DoF. After partial transpose of the  resulting reduced density matrix we compute
\begin{align} \label{ne}
\mathcal{N}=\sum_i \frac{(|\lambda_i|-\lambda_i)}{2},
\end{align}
where $\lambda_i$'s are the eigenvalues of the partial transposed reduced density matrix. Evidently, $\mathcal{N}=0$ implies the reduced system is unentangled.


\begin{figure*}[htbp]
\centering
\includegraphics[scale=0.4]{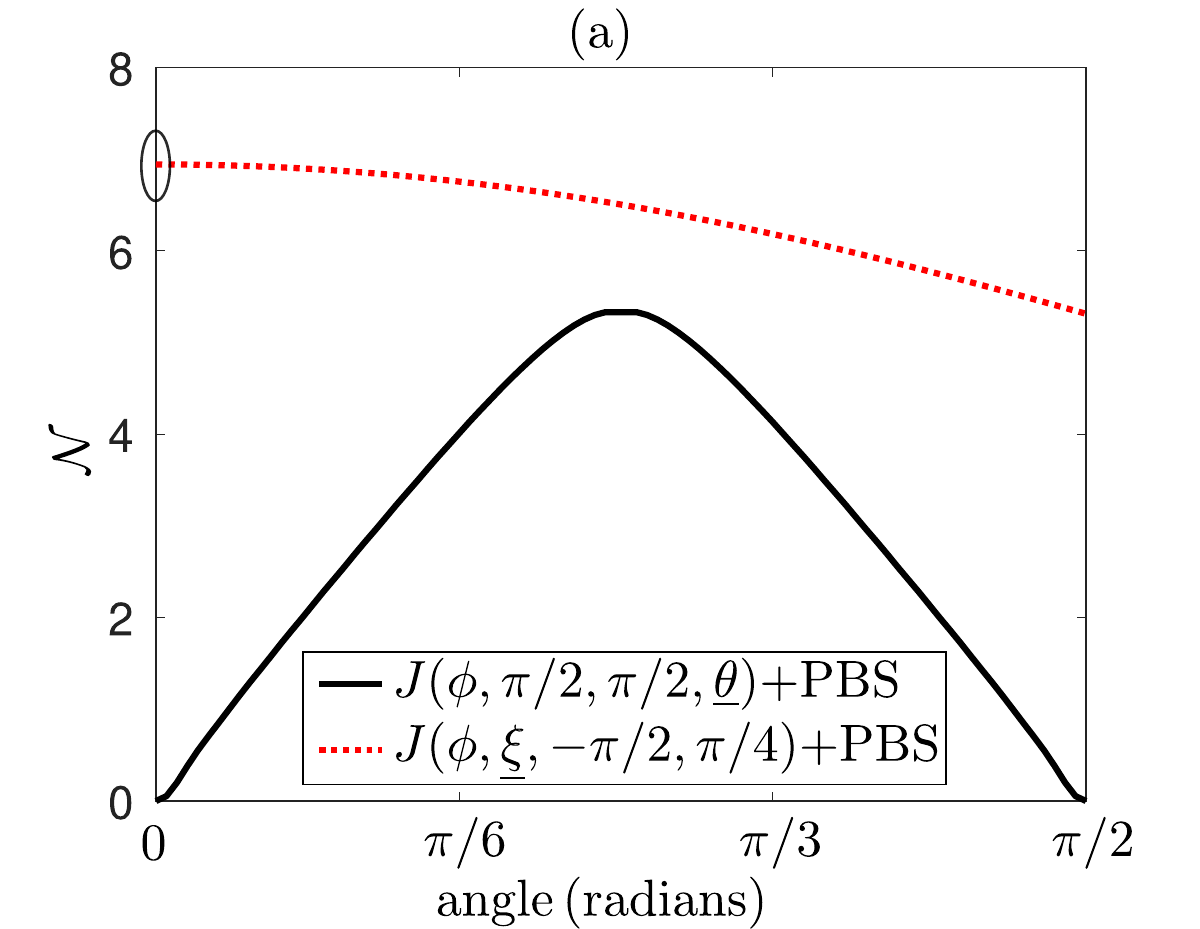}~~\includegraphics[scale=0.4]{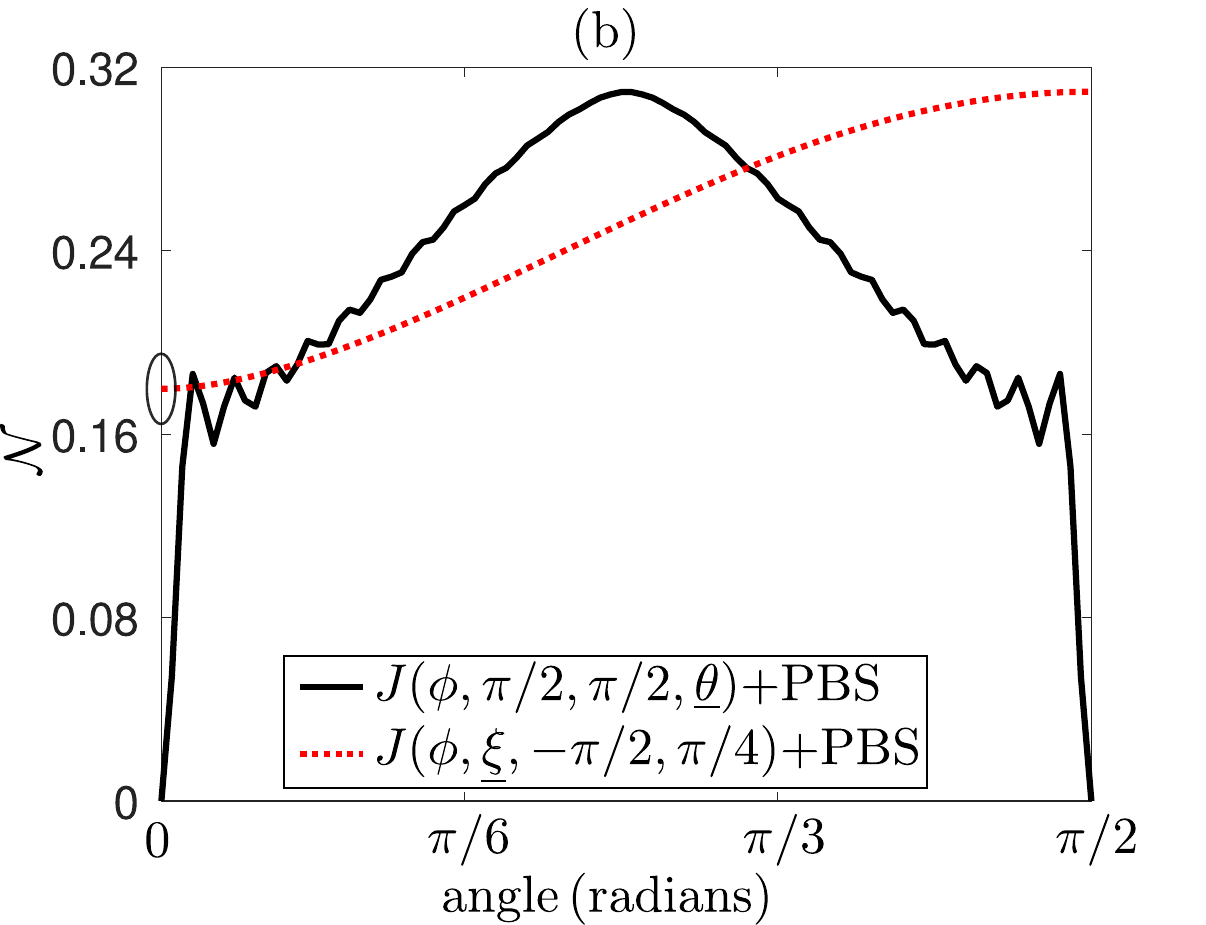}
\caption{Entanglement negativity $\mathcal{N}$\,[see Eq.\,(\ref{ne})] between any two of three DoF plotted against any one of the three $J$-plate parameters $(\xi,\zeta,\theta)$\,[see (\ref{tq12b})] after 25 steps. The initial state was taken to be $[(|H \rangle+|V \rangle)/\sqrt{2}] \otimes |x=0 \rangle \otimes |y=0 \rangle$. In (a), $\mathcal{N}$ between path and OAM DoF is considered, whereas in (b) that between polarization and OAM DoF is considered. In both (a) and (b) the black curve was obtained for $J(\phi,\pi/2,\pi/2,\underline{\theta})$ and PBS combination, where $\underline{\theta}$ denotes that $\theta$ is varied from 0 through $\pi/2$ in steps of $\pi/180$. And the red curve was obtained for $J(\phi,\underline{\xi},-\pi/2,\pi/4)$ and PBS combination, where $\underline{\xi}$ denotes that $\xi$ is varied from 0 through $\pi/2$ in steps of $\pi/180$. Note that the choice $J(\phi, 0, -\pi/2, \pi/4)$ represents a $q$-plate and is encircled in both (a) and (b).}
\label{angle}
\end{figure*}
 
Now we present our numerically simulated results of a single photon passing through the optical setup schematically outlined in the Fig.\,\ref{setup}. The probability distribution of the evolved state after 50 steps for three different $J$-plate parameters $(\xi,\zeta,\theta)$\,[see (\ref{tq12b})] has been shown in the Fig.\,\ref{prob}. Note that $\hat{S}'_{\boldsymbol \sigma,{\rm OAM}}$\,[Eq.\,(\ref{tq13})] with any $(\xi,\zeta,\theta)$\,(or equivalently, $\{|u_1 \rangle, |u_2 \rangle\}$) can be realized using {the $J$-plate + waveplates combination}. Nevertheless, $\hat{S}'_{\boldsymbol \sigma,{\rm OAM}}$ with the choice $(\xi,\zeta,\theta)=(0,-\pi/2,\pi/4)$ can be implemented using a $q$-plate.  In order to demonstrate that the three DoF are entangled, we first trace out the polarization DoF from the density matrix corresponding to $|\Psi_n \rangle$\,(see Eq.\,(\ref{tq10})) and compute the entanglement negativity $\mathcal{N}$\,[see Eq.\,(\ref{ne})] corresponding to the partial transposed reduced density matrix. We then plot $\mathcal{N}$ against increasing number of steps $n$ in Fig.\,\ref{steps}. 

In frame (a) of Fig.\,\ref{steps} we have shown $\mathcal{N}$ between the path and OAM DoF as a function of number of steps. As we increase the number of steps, $\mathcal{N}$ also increases linearly. The value $\mathcal{N}$ can be controlled using the $J$-plate parameters $(\xi,\zeta,\theta)$. If we partial trace any one of the spatial DoF\,(path or OAM DoF), $\mathcal{N}$ corresponding to the reduced density matrix between the polarization and OAM\,(or polarization and path) DoF reaches a steady value as we increase the number of steps\,(see frame (b) of Fig.\,\ref{steps}) for various choices of the $J$-plate parameters $(\xi,\zeta,\theta)$. For instance, for the choice $(\xi,\zeta,\theta)=(0,-\pi/2,\pi/4)$, i.e., a $q$-plate, $\mathcal{N}$ between polarization and OAM\,(or path) DoF reaches a steady state value $0.17927$ after 25 steps, provided we begin with an initial state $[(|H \rangle+|V \rangle)/\sqrt{2}] \otimes |x=0 \rangle \otimes |y=0 \rangle$. 

We now demonstrate how the $J$-plate parameters $(\xi,\zeta,\theta)$, beginning with an initial state, can be used to control the amount of entanglement between three DoF\,: polarization, path, and OAM. In other words, we demonstrate numerically how $\mathcal{N}$ can be controlled by tuning the $J$-plate parameters. In Fig.\,\ref{angle} we have shown how the negativity $\mathcal{N}$ between any two DoF varies with respect to the $J$-plate parameters $(\xi,\zeta,\theta)$. Here, we have allowed one of the three $J$-plate parameters $(\xi,\zeta,\theta)$ to vary while keeping the remaining two constant and plotted the respective $\mathcal{N}$ values. Furthermore, $\mathcal{N}$ between any two DoF has been computed after evolving the quantum walk for 25 steps. In frame (a) of Fig.\,\ref{angle} $\mathcal{N}$ between path and OAM DoF has been plotted as a function of one of the three $J$-plate parameters. And in frame (b) of Fig.\,\ref{angle} $\mathcal{N}$ between polarization and OAM\,(or path) DoF has been obtained as a function of the same. By keeping $\xi=\zeta=\pi/2$ and varying $\theta$ in steps of $\pi/180$ from 0 through $\pi/2$, we obtain a black curve as shown in Fig.\,\ref{angle}. Likewise, keeping $\zeta=-\pi/2$ and $\theta=\pi/4$ and varying $\xi$ in steps of $\pi/180$ from 0 through $\pi/2$, we obtain a red curve as shown in Fig.\,\ref{angle}. The entanglement negativity $\mathcal{N}$ corresponding to the $q$-plate is encircled in both (a) and (b).

\section{Conclusion} 
\label{co}

To summarize, we have proposed a passive optical setup -- using {$J$-plate + waveplates combinations} or $q$-plates, and PBS -- to hyperentangle an incoming single photon in polarization, path, and OAM DoF. We have shown that this optical setup can be efficiently used to simulate the 2D DTQW with variable evolution parameters (modified Pauli walk) without explicitly using a quantum coin operation. The evolved state has been numerically shown to be hyperentangled in polarization, path, and OAM DoF. The amount of entanglement between any two of the three DoF has been computed using entanglement negativity. It was observed that the entanglement negativity increased linearly between the path and OAM DoF, whereas the same between the polarization and path\,(or OAM) DoF remained constant after few number of steps due to the bound on the dimension of the coin space. The amount of entanglement between any two of the three DoF and the hyperentanglement in the system can be controlled by varying the $J$-plate parameters. Ability to control and engineer the dynamics of quantum walks using optical components can also play an important role in realization of non-Markovian quantum channels\,\cite{naikoo2020} and study of open quantum systems. 

\section*{Acknowledgment}
Yasir would like to thank Abhaya S. Hegde for useful discussions on hyperentangled states. Yasir and CMC acknowledge the support from the Office of Principal Scientific Advisor to Government of India, project no. Prn.SA/QSim/2020 and Interdisciplinary Cyber Physical Systems (ICPS) program of the Department of Science and Technology, India, Grant No.: DST/ICPS/QuST/Theme-1/2019/1 for the support.

\appendix

\section{Realization of \texorpdfstring{$J(\phi)$}{} and \texorpdfstring{$\tilde{J}(\phi)$}{}} 
\label{ap}

{In this Appendix we explain how both $J(\phi)$\,[Eq.\,(\ref{tq12a})] and $\tilde{J}(\phi)$\,[Eq.\,(\ref{tq16})] can be realized using a $J$-plate and variable\,(as well as fixed) waveplates. First we note that a $J$-plate can be represented using a Jones matrix\,\cite{devlin2017}
\begin{align} \label{ap1}
\mathbf{J}(\delta_x,\delta_y,\vartheta) &= R_{-\vartheta}
\begin{bmatrix}
e^{i\delta_x} & 0 \\
0 & e^{i\delta_y}
\end{bmatrix} R_\vartheta \nonumber \\
&= \begin{bmatrix}
e^{i\delta_x} \cos^2 \vartheta + e^{i\delta_y} \sin^2 \vartheta & \sin 2\vartheta (-e^{i\delta_x} + e^{i\delta_y})/2 \\
\sin 2\vartheta (-e^{i\delta_x} + e^{i\delta_y})/2 & e^{i\delta_y} \cos^2 \vartheta + e^{i\delta_x} \sin^2 \vartheta
\end{bmatrix},
\end{align}
where
\begin{align} \label{ap2}
R_\vartheta = \begin{bmatrix}
\cos \vartheta & -\sin \vartheta \\
\sin \vartheta & \cos \vartheta
\end{bmatrix}.
\end{align}
In Eq.\,(\ref{ap1}) $e^{i\delta_x}$ and $e^{i\delta_y}$ represent the phase shifts provided by the $J$-plate when $x$- and $y$-polarized light fields pass through it at a point $(x',y')$ in the transverse plane, respectively. Also, $R_\vartheta$ denotes the rotation of the `phase shifter', namely, the diagonal matrix ${\rm diag}\,(e^{i\delta_x}, e^{i\delta_y})$, through an angle $\vartheta$ at a point $(x',y')$ in the transverse plane. It should be kept in mind that the parameters $\delta_x$, $\delta_y$, and $\vartheta$ can be independently controlled and are functions of the point $(x',y')$ in the transverse plane.}

{Now to realize $\tilde{J}(\phi)$\,[Eq.\,(\ref{tq16})] using a $J$-plate, we first parameterize $|u_1 \rangle$ and $|u_2 \rangle$ as
\begin{align} \label{ap3}
|u_1 \rangle = \begin{bmatrix}
\cos \chi \\
e^{i\delta} \sin \chi
\end{bmatrix}, \,\,\,{\rm and} \,\,\,
|u_2 \rangle = \begin{bmatrix}
-\sin \chi \\
e^{i\delta} \cos \chi
\end{bmatrix},
\end{align}
where $0 \leq \chi \leq 2\pi$ and $0 \leq \delta < 2\pi$. With this, we can decompose $\tilde{J}(\phi)$ as
\begin{align} \label{ap4}
\tilde{J}(\phi) = e^{i(\pi-\delta) \sigma_3/2} \, \mathbf{J} (-\phi, \phi, \chi) \, e^{-i(\pi-\delta) \sigma_3/2},
\end{align}
where $\sigma_3={\rm diag}\,(1,-1)$. While $\mathbf{J} (-\phi, \phi, \chi)$ is realized using a $J$-plate\,[see Eq.\,(\ref{ap1})], both first and third matrices on the RHS of Eq.\,(\ref{ap4}) are realized using one variable waveplate each.
In the case of $J(\phi)$\,[Eq.\,(\ref{tq12a})], it can be verified that
\begin{align} \label{ap5}
J(\phi) = e^{-i\delta \sigma_3/2} (i\sigma_1) \, \mathbf{J} (-\phi, \phi+\pi, \chi) \, e^{-i(\pi-\delta) \sigma_3/2}.
\end{align}
Here, both $e^{-i\delta \sigma_3/2}$ and $e^{-i(\pi-\delta) \sigma_3/2}$ can be realized using one variable waveplate each, and $i\sigma_1$ can be realized using $H_{\pi/4}$, namely, HWP rotated through an angle $\pi/4$\,\cite{simon90}. 
Hence we require a $J$-plate, 1 HWP, and 2 variable waveplates to realize $J(\phi)$.
With these, we conclude that $\tilde{J}(\phi)$\,[see Eq.\,(\ref{tq16})] requires a $J$-plate and 2 variable waveplates. On the other hand, $J(\phi)$\,[see Eq.\,(\ref{tq12a})] requires an additional half waveplate.
}



\twocolumngrid

\end{document}